\documentclass[superscriptaddress,showpacs,preprintnumbers,amsmath,amssymb]{revtex4}
\usepackage{graphicx}
\usepackage{dcolumn}
\usepackage{bm}
\usepackage{latexsym}
\begin{document}
\title{A two-state model for helicase translocation and unwinding of
  nucleic acids}

\author{Ashok Garai{\footnote{E-mail: garai@iitk.ac.in}}}
\affiliation{Department of Physics, Indian Institute of Technology,
  Kanpur 208016, India.}

\author{Debashish Chowdhury{\footnote{E-mail: debch@iitk.ac.in}}}
\affiliation{Department of Physics, Indian Institute of Technology,
  Kanpur 208016, India; and\\
  Max-Planck Institute for Physics of Complex Systems, N\"othnitzer
  Strasse 38, D-01187 Dresden, Germany.}

\author{M. D. Betterton{\footnote{E-mail: mdb@colorado.edu}}}
\affiliation{Physics Department, University of Colorado, Boulder, CO
  80309-0390, U.S.A.}

\date{\today}%
\begin{abstract}
  Helicases are molecular motors that unwind double-stranded nucleic
  acids (dsNA), such as DNA and RNA). Typically a helicase
  translocates along one of the NA single strands while unwinding and
  uses adenosine triphosphate (ATP) hydrolysis as an energy source.
  Here we model of a helicase motor that can switch between two
  states, which could represent two different points in the ATP
  hydrolysis cycle.  Our model is an extension of the earlier
  Betterton-J\"ulicher model of helicases to incorporate switching
  between two states.  The main predictions of the model are the speed
  of unwinding of the dsNA and fluctuations around the average
  unwinding velocity.  Motivated by a recent claim that the NS3
  helicase of Hepatitis C virus follows a flashing ratchet mechanism,
  we have compared the experimental results for the NS3 helicase with
  a special limit of our model which corresponds to the flashing
  ratchet scenario. Our model accounts for one key feature of the
  experimental data on NS3 helicase. However, contradictory
  observations in experiments carried out under different conditions
  limit the ability to compare the model to experiments.
\end{abstract}
\maketitle

Helicases are enzymes that unwind double-stranded nucleic acids (dsNA)
\cite{alberts02}. Helicase proteins typically translocate along one of
the single strands and perform mechanical work while consuming
chemical energy (usually supplied by the hydrolysis of ATP).
Therefore, these NA translocases are molecular motors
\cite{schliwa03,mavroidis04} which share common features with
cytoskeletal molecular motors \cite{lohman98,howard01}.

All helicases undergo a biochemical cycle which typically involves ATP
binding, ATP hydrolysis, and release of the hydrolysis products
adenosine diphosphate (ADP) and and inorganic phosphate (P$_i$).  An
important question in the study of helicase mechanisms is to
understand how the ATP hydrolysis cycle is coupled to the binding
state and the motion of the helicase \cite{lohman96,delagoutte02}.
Helicases may exhibit changes in helicase/NA binding affinity when the
helicase is bound to ATP, ADP/P$_i$, or neither; coordination of
hydrolysis between different helicase subunits, and conformational
changes in the helicase triggered by different steps in the hydrolysis
cycle.  Some helicases form hexamers (which include six ATPase
domains), while others are members of the non-hexameric (dimeric or
monomeric) group; different types of mechanochemical cycle have been
suggested for the different structural classes
\cite{lohman96,patel00}.  In all cases, one seeks to explain how the
helicase coordinates NA binding and hydrolysis to move along
single-stranded NA and unwind double-stranded NA.

Here we develop a generic model of a helicase that switches between
two biochemical states while translocating on ssNA. This is a
simplified representation of the different states of the helicase
during the ATP hydrolysis cycle. The model may be generally applicable
to helicases for which the transition between two states is the key
feature of the motion. In other words, this model should be a good
approximation for helicases with more than two biochemical states if
one transition is far slower than the others. We incorporate such a
two-state picture by extending the original Betterton-J\"ulicher (BJ)
model~\cite{betterton03,betterton05a,betterton05b} of NA
helicases~\cite{note1}.

Our work is also connected to two-state models that have been used
extensively for a variety of molecular motors
\cite{julicher97,nishi05,zeldo05,fisher07}.  Under a mean-field
approximation, such models can be easily solved when periodic boundary
conditions are imposed. However, the problem is usually more difficult
with open boundary conditions. The model for helicase motion is even
more complex because the position of one boundary (i.e., the ssNA-dsNA
junction) varies randomly with time. Thus our work is also an
extension of previous work on two-state models to the more difficult
case of a fluctuating boundary.

The two-state model developed here is consistent with the observation
that binding and hydrolysis of ATP can modulate the affinity of a
helicase for the nucleic-acid track
\cite{bjornson96,kim98,velankar99}.  The flashing-ratchet mechanism
suggested qualitatively for the hepatitis C virus non-structural
protein 3 (HCV NS3) helicase \cite{levin03,levin05} can be captured by
a special case of the generic model proposed here. In the
flashing-ratchet \cite{julicher97} picture, the motor protein switches
between two states: one where the protein is tightly bound to the
track, and another where the motor is weakly bound and can diffuse
along the track. In this paper we make quantitative comparisons
between our theoretical predictions for a passive helicase which
follows the flashing-ratchet mechanism, and the experimental data for
NS3 helicase.

In section \ref{sec:model} we describe the ingredients of the model:
the helicase, which can switch between two states and translocate on
ssNA, and the fluctuating NA ss-ds junction.  In section
\ref{sec:sseqns} we calculate the single-strand translocation rate of
the helicase. Section \ref{sec:eqns} contains the model equations for
double-strand unwinding, the transformation of the equations using
midpoint and difference variables, and the general solutions for the
velocity and diffusion coefficient. We describe the results for a
hard-wall interaction between the helicase and junction in section
\ref{sec:soln}.  Using rate constants estimated from experiments on
NS3 helicase, in section \ref{sec:results} we specialize to the
flashing-ratchet scenario and make predictions specific to NS3.  In
section \ref{sec:disc} we summarize our results.

\section{The model}
\label{sec:model}

Here we develop a physical model for a helicase that moves on ssNA
while cycling between two chemical states (labeled 1 and 2). Levin et
al.~suggested such a two-state model for NS3 helicase motion
\cite{levin03,levin05}.  In this paper, we first consider a general
two-state model, and later focus on the specific flashing-ratchet
picture.

\begin{figure}
\begin{center}
\includegraphics[width=3in]{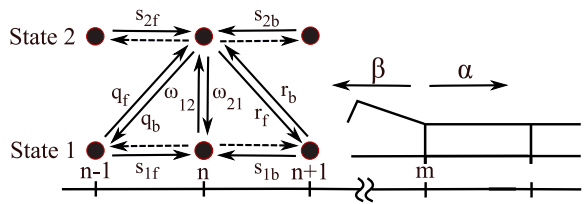}
\end{center}
\caption{Schematic of the model. The protein can exist in either of
  two chemical states (labeled 1 and 2) at each lattice site (labeled
  $n$). Sliding transitions (where $n$ changes but the chemical 
  state does not) occur at rate $s_{1f}$, etc., depending on the 
  state and whether the transition is forward (toward increasing $n$) 
  or backward (toward decreasing $n$). Chemical transitions (where 
  the chemical state changes but $n$ does not) occur at rates 
  $\omega_{12}$ (for the transition from 1 to 2) and $\omega_{21}$ 
  (for the transition from 2 to 1). Coupled transitions, where both 
  the chemical state and $n$ change, occur at rates $r_f$ (for 
  the transition from 2 to 1 coupled to forward motion),
  $r_b$ (for the transition from 1 to 2 coupled to backward motion),
  $q_f$ (for the transition from 1 to 2 coupled to forward motion),
  and $q_b$ (for the transition from 2 to 1 coupled to backward
  motion).  The nucleic acid single strand-double strand junction is
  at site $m$. The junction moves toward increasing $m$ when the NA
  opens by one base (rate $\alpha$) and toward decreasing $m$ when the
  NA closes (rate $\beta$). }
\label{fig-model}
\end{figure}

In the traditional continuous models of Brownian ratchets, one first
writes a Fokker-Planck equation. We use a discrete model, so our
approach is based on master equations. The discrete approach can be
useful when comparing to experiments. In the Fokker-Planck approach,
one needs the explicit functional form of the fluctuating potential,
which has not been measured for any real motor.  In the discrete
model, we bypass this difficulty by capturing the motor mechanism
through a choice of rate constants (or transition probabilities), many
of which can be obtained from experiments (see section
\ref{sec:results}).

In the discrete model, we represent the ssNA by a one-dimensional
lattice where each site corresponds to a single base. We label each
site by the integer index $i$. As in the BJ model \cite{betterton03},
we neglect the sequence inhomogeneity of the ssNA (in principle, the
model can be extended to capture this feature, which may be important
in some limits \cite{kafri04}).  The position of the helicase is
denoted by the integer $n$. Most helicases have a fixed direction of
translocation, either $3'$ to $5'$ or $5'$ to $3'$ along the
left-right asymmetric ssNA \cite{lohman96}. In our model the helicase
translocates toward increasing $n$ (from left to right in
fig.~\ref{fig-model}). At any spatial position $n$, the helicase can
be either in biochemical state 1 or 2. 

The model is fully described by the allowed transitions between states
and the corresponding reaction rates. In general, we could have all
transitions sketched in fig. \ref{fig-model}.  Helicase ``sliding''
corresponds to transitions along the ssNA without a change in
biochemical state of the protein. In state 1, these sliding
transitions occur at rate $s_{1f}$ (for increasing $n$) and $s_{1b}$
(for decreasing $n$).  When the helicase is in state 2, the
forward/backward sliding rates are $s_{2f}$ and $s_{2b}$. Physically,
these transitions occur because of Brownian motion of the protein,
decoupled from any biochemical state change.  

The helicase can undergo ``chemical'' transitions which correspond to
a change in biochemical state without physical translocation along the
ssNA. At fixed $n$, the rate of transition from state 1 to 2 occurs at
rate $\omega_{12}$, while the reverse transition occurs at rate
$\omega_{21}$.  Finally, ``coupled'' mechanochemical transitions are those
where a change of biochemical state and physical translocation occur
together. If the helicase is located at $n$ and is in state 2, then it
can make a transition to state 1 while moving forward to site $n+1$ at
rate $r_{f}$; the corresponding reverse rate is $r_b$. The transition
of the helicase from state 1 to 2 while moving forward from $n$ to
$n+1$ occurs at rate $q_f$; the corresponding reverse rate is $q_b$.

If any of these reactions is coupled to ATP hydrolysis, then the
forward/reverse transitions may be out of equilibrium and break the
detailed balance relation.  The Levin et al.~model of HCV NS3 helicase
suggests that ATP binding is required to remove the helicase from the
tightly bound state \cite{levin03,levin05}, implying that the $1 \to
2$ transition at rate $\omega_{12}$ is determined by the ATP
concentration. In the Levin et al.~flashing-ratchet model, ATP
hydrolysis and product release is coupled to the translocation and
chemical transition back to state $1$, which in our representation
means that rates $\omega_{21}$ and $r_f$ would be coupled to ATP
hydrolysis and would therefore be out of equilibrium (see section
\ref{sec:results}).

The junction between ssNA and dsNA is labeled by $m$ (see
fig.~\ref{fig-model}). The dsNA opens and closes due to thermal
fluctuations. When the helicase and junction are far apart, the
opening rate is $\alpha$ and the closing rate $\beta$. We assume that
these rates are independent of the NA base sequence and that the only
fluctuations are those for which the NA opens or closes at the ss-ds
fork.  Following the BJ model \cite{betterton03}, we neglect the
possibility of any jump $>1$ bp in the position of the ssNA-dsNA
junction. However, this approximation is justified because, at the
temperatures of our interest (i.e., sufficiently below the melting
temperature of the dsDNA) the spontaneous formation of bubbles is
rare.  Since the NA breathing results from thermal fluctuations, the
rates $\alpha$ and $\beta$ satisfy detailed balance: $
\frac{\alpha}{\beta}=e^{-\Delta G}$, where $\Delta G$ is the free
energy of one base-pair bond in units of $kT$.

The main quantity of interest is the speed of unwinding of dsNA by a
helicase. We derive an analytical expression for the unwinding
velocity. We compare the predicted velocity with the corresponding
experimental data for a specific helicase, NS3 helicase of hepatitis C
virus. Although we also derive an analytical expression for the
diffusion constant of the helicase, we do not compare it with
experimental data for any specific helicase.

In this work we analyze passive unwinding, which is equivalent to a
hard-wall interaction potential in the BJ model \cite{betterton05a}.
In passive unwinding, the helicase acts as a block to NA closing when
adjacent to the junction. The protein moves forward only when thermal
fluctuations open a basepair at the NA ss-ds junction. This means that
when the helicase and junction are adjacent ($j=1$), the helicase
cannot hop forward (all helicase forward rates, $s_{1f}(j=1)$,
$s_{2f}(j=1)$, $r_f(j=1)$, and $q_f(j=1)$, are zero) and the NA cannot
close ($\beta(j=1)=0$). Otherwise, the rates are unaffected by the
helicase-junction interaction.

\section{Single-strand translocation}
\label{sec:sseqns}

In order to motivate our approach, we first formulate the equations
for a helicase sufficiently far from the ssNA-dsNA junction so that it
translocates on ssNA without any dsNA unwinding activity.  Let ${\cal
  P}_{\mu}(n,t)$ denote the probability that, at time $t$, the
helicase is located at site $n$ and is in the chemical state $\mu$. We
will drop the reference to the time dependence of ${\cal P}_{\mu}(n)$.
The master equations governing the time evolution of ${\cal
  P}_{\mu}(n)$ are
\begin{eqnarray}
  \frac{d{\cal P}_{1}(n)}{dt} &=&
  -(\omega_{12}+s_{1f}+s_{1b}+q_f+r_b){\cal P}_{1}(n) + 
  s_{1f}{\cal P}_{1}(n-1) + r_f {\cal P}_{2}(n-1)
  \nonumber \\    
  &+&s_{1b}{\cal P}_{1}(n+1) +q_b {\cal P}_{2}(n+1)
  + \omega_{21}{\cal P}_{2}(n),
\label{eq-master1ss}
\end{eqnarray}
and
\begin{eqnarray}
  \frac{d{\cal P}_{2}(n)}{dt} &=&
  -(\omega_{21}+s_{2f}+s_{2b}+r_f+q_b){\cal P}_{2}(n) + 
  s_{2f}{\cal P}_{2}(n-1) + q_f {\cal P}_{1}(n-1)
  \nonumber \\    
  &+&s_{2b}{\cal P}_{2}(n+1) +r_b {\cal P}_{1}(n+1)
  + \omega_{12}{\cal P}_{1}(n).
\label{eq-master2ss}
\end{eqnarray}
Summing these equations, we find the total probability ${\cal P}(n)={\cal
  P}_1(n)+{\cal P}_2(n)$ satisfies
\begin{eqnarray}
  \frac{d{\cal P}(n)}{dt} &=&
  -(s_{1f}+s_{1b}+q_f+r_b){\cal P}_{1}(n)
  -(s_{2f}+s_{2b}+r_f+q_b){\cal P}_{2}(n) + 
  (s_{1f}+q_f){\cal P}_{1}(n-1) + (s_{2f}+r_f) {\cal P}_{2}(n-1)
  \nonumber \\    
  &+&(s_{1b}+r_b){\cal P}_{1}(n+1) +(s_{2b}+q_b) {\cal P}_{2}(n+1).
\label{eq-masterss}
\end{eqnarray}
These equations have a translationally invariant steady-state solution
where ${\cal P}_{\mu}(n)$ is independent of $n$. In this case, we
expect that the probability in state 2 is a multiple of the
probability in state 1:
\begin{equation}
{\cal P}_{2}(n)=\sigma  {\cal P}_{1}(n),
\label{eq-propss}
\end{equation}
which means that ${\cal P}(n)=(1+\sigma){\cal P}_1(n)$. 

In this case, the master equation for the total probability can be 
written as a hopping model with effective rates $k_f$ for forward 
transitions and $k_b$ for backward transitions. At steady state,
\begin{eqnarray}
  0 &=& k_f {\cal P}(n-1) - (k_f + k_b) {\cal
    P}(n) + k_b {\cal P}(n+1),
\label{eq-hop}
\end{eqnarray}
where 
\begin{eqnarray}
 k_f &=& \frac{s_{1f}+q_f+\sigma(s_{2f}+r_f)}{1+\sigma}, \label{eq-kf}\\
 k_b &=& \frac{s_{1b}+r_b+\sigma(s_{2b}+q_b)}{1+\sigma}.
  \label{eq-kb}
\end{eqnarray}
and the expression 
\begin{equation}
\sigma=
\frac{\omega_{12}+q_f+r_b}{r_f+q_b+\omega_{21}}.
\end{equation}
has been obtained from eqn.~\eqref{eq-master1ss} at steady state,
assuming translational invariance.  The mean single-strand
translocation velocity is $v_{ss}=k_f-k_b$. 

\section{Double-strand unwinding: Model equations}
\label{sec:eqns}
In this section we extend the formulation of the preceding section by
incorporating helicase-catalyzed dsNA unwinding.  Let ${\cal
  P}_{\mu}(n,m;t)$ denote the probability that, at time $t$, the
helicase is at located at $n$ and is in the chemical state $\mu$,
while the ss-ds junction is at $m$.  We will drop the reference to the
time dependence of ${\cal P}_{\mu}(n,m)$. The master equations
governing the time evolution of ${\cal P}_{\mu}(n,m)$ are given by
\begin{eqnarray}
  \frac{d{\cal P}_{1}(n,m)}{dt} &=& -(\alpha + \beta+ 
  \omega_{12}+s_{1f}+s_{1b}+q_f+r_b){\cal P}_{1}(n,m) +
  s_{1f}{\cal P}_{1}(n-1,m) + r_f {\cal P}_{2}(n-1,m)
  \nonumber \\    
  &+&s_{1b}{\cal P}_{1}(n+1,m) +q_b {\cal P}_{2}(n+1,m)
  + \omega_{21}{\cal P}_{2}(n,m)+\alpha {\cal P}_{1}(n,m-1)
  +\beta P_{1}(n,m+1) ~~(m > n).  
\label{eq-master1}
\end{eqnarray}
and
\begin{eqnarray}
  \frac{d{\cal P}_{2}(n,m)}{dt} &=& -(\alpha + \beta+ 
  \omega_{21}+s_{2f}+s_{2b}+r_f+q_b){\cal P}_{2}(n,m) +
  s_{2f}{\cal P}_{2}(n-1,m) + q_f {\cal P}_{1}(n-1,m)
  \nonumber \\    
  &+&s_{2b}{\cal P}_{2}(n+1,m) +r_b {\cal P}_{1}(n+1,m)
  + \omega_{12}{\cal P}_{1}(n,m)+\alpha {\cal P}_{2}(n,m-1)
  +\beta {\cal P}_{2}(n,m+1) ~~(m > n).  
\label{eq-master2}
\end{eqnarray}
Note that the rates depend on the separation $m-n$; this notation is
omitted for clarity.  We assume the interaction
potential is the same for both chemical states, so that the
position-dependent NA opening and closing rates $\alpha$ and $\beta$
are independent of the chemical state.

Next we change variables to work with the difference $j = m-n$ and
midpoint $l= 2 l^{'}=m+n$ positions of the helicase-junction
complex. Rewriting eqns.~\eqref{eq-master1} and \eqref{eq-master2}
we have
\begin{eqnarray}
  \frac{d{\cal P}_{1}(j,l)}{dt} &=& -(\alpha + \beta+ 
  \omega_{12}+s_{1f}+s_{1b}+q_f+r_b){\cal P}_{1}(j,l) +
  s_{1f}{\cal P}_{1}(j+1,l-1) + r_f {\cal P}_{2}(j+1,l-1)
  \nonumber \\    
  &+&s_{1b}{\cal P}_{1}(j-1,l+1) +q_b {\cal P}_{2}(j-1,l+1)
  + \omega_{21}{\cal P}_{2}(j,l)+\alpha {\cal P}_{1}(j-1,l-1)
  +\beta {\cal P}_{1}(j+1,l+1) \nonumber \\ 
  && ~~~~~~~~~~~~~~~~~~~~~~~~~~~~~~~~~~~~~~~~~~~~~~~~~~~~~~~~~~~~~~~~~~~~~~~~~~~~~~~~~~~~~~~~~~~~(j > 0).  
\label{eq-master1j}
\end{eqnarray}
and
\begin{eqnarray}
  \frac{d{\cal P}_{2}(j,l)}{dt} &=& -(\alpha + \beta+ 
  \omega_{21}+s_{2f}+s_{2b}+r_f+q_b){\cal P}_{2}(j,l) +
  s_{2f}{\cal P}_{2}(j+1,l-1) + q_f {\cal P}_{1}(j+1,l-1)
  \nonumber \\    
  &+&s_{2b}{\cal P}_{2}(j-1,l+1) +r_b {\cal P}_{1}(j-1,l+1)
  + \omega_{12}{\cal P}_{1}(j,l)+\alpha {\cal P}_{2}(j-1,l-1)
  +\beta {\cal P}_2(j+1,l+1) \nonumber \\ 
  && ~~~~~~~~~~~~~~~~~~~~~~~~~~~~~~~~~~~~~~~~~~~~~~~~~~~~~~~~~~~~~~~~~~~~~~~~~~~~~~~~~~~~~~~~~~~~(j > 0).
\label{eq-master2j}
\end{eqnarray}
Again, the rates vary with $j$. However, the rates are independent of
$l$, so we can sum over the position of the complex center of mass:
\begin{eqnarray}
P_{1}(j) = \sum_{l} {\cal P}_{1}(j,l) \nonumber \\
P_{2}(j) = \sum_{l}{\cal P}_{2}(j,l)
\label{eq-defp}
\end{eqnarray}
Applying the sum over $l$ to eqns.\~(\ref{eq-master1j}) and
(\ref{eq-master2j}) we find
\begin{eqnarray}
  \frac{d P_{1}(j)}{dt} &=& -(\alpha + \beta+ 
  \omega_{12}+s_{1f}+s_{1b}+q_f+r_b) P_{1}(j) +
  (s_{1f}+\beta) P_{1}(j+1) + r_f  P_{2}(j+1)
  \nonumber \\    
  &+&(s_{1b}+\alpha) P_{1}(j-1) +q_b  P_{2}(j-1)
  + \omega_{21} P_{2}(j) .  
\label{eq-dif1}
\end{eqnarray}
and
\begin{eqnarray}
  \frac{d P_{2}(j)}{dt} &=& -(\alpha + \beta+ 
  \omega_{21}+s_{2f}+s_{2b}+r_f+q_b) P_{2}(j) +
  (s_{2f}+\beta) P_{2}(j+1) + q_f  P_{1}(j+1)
  \nonumber \\    
  &+&(s_{2b}+\alpha) P_{2}(j-1) +r_b  P_{1}(j-1)
  + \omega_{12} P_{1}(j).
\label{eq-dif2}
\end{eqnarray}
We consider the total probability by summing eqns.~\eqref{eq-dif1} and
\eqref{eq-dif2}. Defining the total probability current
\begin{equation}
  I(j)= \alpha P(j) -\beta P(j+1) + (s_{1b}+r_b) P_1(j)+(s_{2b}+q_b) P_2(j)
 -(s_{1f}+q_f) P_1(j+1) - (s_{2f}+r_f) P_2  (j+1),
\label{eq-curr}  
\end{equation}
the total probability satisfies
\begin{equation}
  \frac{d P(j)}{dt} =-I(j)+I(j-1).
\label{eq-dif}
\end{equation}

At steady state $P(j)$ is time independent, so $I(j) = I(j-1)$.
Further, since $U(j) \rightarrow \infty$ as $j\rightarrow -\infty$,
this constant probability flux must be zero, i.e.,
$I(j) = 0$ for all $j$. 

Adding the two eqns.~(\ref{eq-master1j}) and (\ref{eq-master2j}) and
defining ${\cal P}(j,l) = {\cal P}_{1}(j,l) + {\cal P}_{2}(j,l)$, we
get
\begin{eqnarray}
  \frac{d{\cal P}(j,l)}{dt} &=& -(\alpha + \beta) {\cal P}(j,l) + \alpha
  {\cal P}(j-1,l-1) + \beta P(j+1,l+1) 
  +( s_{1f} + q_f){\cal P}_{1}(j+1,l-1) \nonumber \\
  &+& (r_f+s_{2f}) {\cal P}_{2}(j+1,l-1) +(s_{1b}+r_b){\cal
    P}_{1}(j-1,l+1) +(q_b+s_{2b}) {\cal P}_{2}(j-1,l+1)  \nonumber \\ 
  &+& \omega_{21}{\cal P}_{2}(j,l)+   \omega_{12}{\cal
    P}_{1}(j,l) -(\omega_{12}+s_{1f}+s_{1b}+q_f+r_b) {\cal P}_{1}(j,l)
  \nonumber \\
  &-&(\omega_{21}+s_{2f}+s_{2b}+r_f+q_b){\cal P}_{2}(j,l).
\label{eq-masterjl}
\end{eqnarray}
The probability distribution in $l$ at time 
$t$ is
\begin{equation}
  \Pi(l;t) = \sum_{j} {\cal P}(j,l;t)
\label{eq-defpi}
\end{equation}
Note that, by definition, $\Pi(l;t)$ is independent of the chemical
state of the helicase For times much longer than the relaxation time
of the difference variable $j$, we can assume
\begin{equation}
{\cal P}_{\mu}(j,l) =  P_{\mu}(j) ~\Pi(l) \quad (\mu = 1 ~{\rm or} ~2)
\end{equation}
Starting from the eqn.~(\ref{eq-masterjl}), one can derive
\begin{equation}
\frac{d\Pi(l)}{dt} = u \Pi(l-1) - (u+w) \Pi(l) +  w \Pi(l+1)
\label{eq-pieqn}
\end{equation}
where 
\begin{eqnarray}
  u = \sum_{j} \alpha  P(j) + (s_{1f}+q_f) { P}_{1}(j) +
    (s_{2f}+r_f ) { P}_{2}(j),
\label{eq-u}
\end{eqnarray}
and 
\begin{eqnarray}
w = \sum_{j}  \beta { P}(j) + (s_{1b}+r_b) { P}_{1}(j) +
  (s_{2b}+q_b)  {  P}_{2}(j).
\label{eq-w}
\end{eqnarray}
Thus the motion of the helicase-junction complex is a combination of
drift and diffusion. Note that in the special case $u = w$ the drift
vanishes and the dynamics of $l$ becomes purely diffusive.

As in ref.~\cite{betterton05a}, the average speed of unwinding is $v =
\frac{1}{2}(u-w)$, or
\begin{equation}
  v=\frac{1}{2} \sum_{j}  (\alpha-\beta)  P(j) +
  (s_{1f}+q_f-s_{1b}-r_b) P_{1}(j) + 
  (s_{2f}+r_f -s_{2b}-q_b) P_{2}(j).
\label{eq-v}
\end{equation}
Similarly, the diffusion coefficient is $D =\frac{1}{4} (u+w)$, which is
\begin{equation}
D=\frac{1}{4} \sum_{j}  (\alpha+\beta)  P(j) +
  (s_{1f}+q_f+s_{1b}+r_b) P_{1}(j) + 
    (s_{2f}+r_f +s_{2b}+q_b) P_{2}(j).
\label{eq-d}
\end{equation}
Note that if the sliding transitions represent unbiased diffusion,
then the forward and backward rates  $s_{\mu f}$ and $s_{\mu b}$ are
equal. Then the terms involving the sliding rates
drop out from the expression for $v$ but not from that for $D$.

\section{Solution}
\label{sec:soln}

In order to evaluate the expressions for the unwinding velocity and
diffusion coefficient, we must determine $ P_{1}(j)$ and $ P_{2}(j)$
in terms of the rate constants.  Consider the result of summing
eqns.~\eqref{eq-dif1} and \eqref{eq-dif2} over $j$ to determine
equations for the total probability of being in state 1, $P_1$ and the
total probability of being in state 2, $P_2$.  We can write these
equations as
\begin{eqnarray}
  \frac{d P_{1}}{dt} &=& -k_{12} P_{1} + k_{21}
   P_{2},  \\    
  \frac{d P_{2}}{dt} &=& -k_{21} P_{2} + k_{12} P_1,
\label{eq-jsum}
\end{eqnarray}
where the rate constant $k_{12}$ depends on $\omega_{12}$, $q_f$, and
$r_b$, and $k_{21}$ depends on $\omega_{21}$, $r_f$, and $q_b$. The
steady-state solution has $P_2 = k_{12}/k_{21} P_1$.
  
This observation suggests a translationally invariant solution for
$P_1(j)$ and $P_2(j)$ when the rates are constant.  We consider the
case where the relative probability of being in state 1 or 2 is
translationally invariant (independent of $j$). This must occur if the
hopping rates are constant or spatially vary in the same way (for
example, if states 1 and 2 have the same interaction potential with
the dsNA). Since we are primarily interested in a passive helicase
with constant rates, we will focus on this case.  Because of the
translational invariance, the probability in state 2 is a multiple of
the probability in state 1, so that
\begin{equation}
P_{2}(j)=\gamma  P_{1}(j).
\label{eq-prop}
\end{equation}

The zero-current relation requires that
eqn.~\eqref{eq-curr} equal zero, which requires
\begin{equation}
  (\beta+s_{1f}+q_f) P_1(j+1) + (\beta +s_{2f}+r_f) P_2
  (j+1)=(\alpha+s_{1b}+r_b) P_1(j)+(\alpha+s_{2b}+q_b) P_2(j) .
\label{eq-recur}  
\end{equation}
We can plug in to eqn.~\eqref{eq-prop} and solve for the unknown
constant $\gamma$. We can rewrite eqn.~\eqref{eq-recur} as a recursion
relation that relates $P_1(j+1)$ to $P_1(j)$:
\begin{equation}
  \frac{P_1(j+1)}{P_1(j)} = \frac{\alpha (1 +\gamma) +s_{1b}+r_b +\gamma
    (s_{2b}+q_b) }{  \beta(1+\gamma)+s_{1f}+q_f + \gamma
    (s_{2f}+r_f)} = c.  
\label{eq-recur1}  
\end{equation}
Note that $c$ is a function of $\gamma$.  While it is possible to
solve coupled equations for $c$ and $\gamma$ in general, the resulting
expressions are long and not useful for developing intuition. Instead,
we use the approximation relevant for helicases that $\alpha$ and
$\beta$, the opening and closing rates of the NA, are several orders
of magnitude larger than the other rates in the problem  (see
  reference \cite{betterton05a}, where experimental data from
  reference \cite{bonnet98} was used to estimate the opening rate
  $\alpha \sim 10^7$ s$^{-1}$; other rates in the problem are of order
  $10^2$ s$^{-1}$). In this case, eqn.~\eqref{eq-recur1} reduces to
\begin{equation}
  \label{eq-c}
  c\approx \frac{\alpha}{\beta}. 
\end{equation}
Throughout the remainder of this paper, we will use this approximate
value of $c$. Note that because $\alpha$ and $\beta$ are constant, $c$
is also constant and eqn.~\eqref{eq-recur1} shows that $P_1(j)$ has
power-law decay with increasing $j$ (as in the BJ model for a passive
helicase \cite{betterton05a}).

 Using eqns.~\eqref{eq-prop} and \eqref{eq-recur1} in
  eqn.~\eqref{eq-dif1} at steady state, and imposing the requirement
  that $P_1(j)$ cannot vanish for arbitrary $j$, we find a unique
  expression for $\gamma$:
\begin{equation}
  \gamma= \frac{s_{1f}(1-c)-s_{1b}(c^{-1}-1)+r_b+q_f+\omega_{12}}
  {cr_f+c^{-1}q_b+\omega_{21}}. 
\label{eq-gamma}  
\end{equation}

With this result, we can evaluate eqns.~\eqref{eq-v} and \eqref{eq-d}
and express $v$ and $D$ in a fashion analogous to the expressions in
the simpler BJ model:
\begin{equation}
  v = \frac{1}{2} \sum_j {\cal P}_1(j) (a + k^+ - b - k^-), 
\label{eq-vfin}
\end{equation}
\begin{equation}
  D = \frac{1}{4} \sum_j {\cal P}_1(j) (a + k^+ + b + k^-),
\label{eq-dfin}
\end{equation}
if we define the effective rates
\begin{eqnarray}
  a &=& \alpha (1+\gamma), \\
  b &=& \beta (1+\gamma), \\
  k^+ &=&  \gamma (s_{2f}+r_f) + s_{1f}+q_f ,\\
  k^- &=&\gamma  (s_{2b}+q_b)+s_{1b}+r_b.
\end{eqnarray}
Next we evaluate the sums in eqns.~(\ref{eq-vfin}) and (\ref{eq-dfin}),
noting that $P_1(j)=P_1 c^j$ and taking into account that for $j=1$
the rates $k^+$ and $b$ are zero. The result is
\begin{equation}
  v = \frac{c k^+ - k^-}{2(1+\gamma)}, 
\label{vel}
\end{equation}
\begin{equation}
  D = \frac{\alpha}{2} + \frac{c k^+ +k^-}{4(1+\gamma)}
\label{diff}
\end{equation}
Equations (\ref{vel}) and (\ref{diff}) are the main results.

Note that under most conditions the NA opening and closing rate
$\alpha$ is orders of magnitude larger than the other rates, and
therefore $D \approx \alpha/2$.

\section{Comparison with NS3 helicase}
\label{sec:results}

The NS3 helicase of the hepatitis C virus (HCV) is important for HCV
replication, and is therefore a potential drug target \cite{frick07}.
NS3 is also an interesting model helicase because it is the only
currently known helicase capable of unwinding both dsRNA and dsDNA
\cite{kim95,tai96}.  The flashing-ratchet mechanism proposed for NS3
helicase in ref.~\cite{levin05} is a special case of the two-state
model which we have developed in the preceding sections.  In this
section, we first briefly summarize the experimental data on NS3
helicase and their mutually contradictory interpretations which
highlight the current debates in the literature.  Then, we present
analytical results for the special case of our model which captures
the flashing ratchet mechanism. We compare these theoretical
predictions with the corresponding experimental data for NS3 helicase.
The comparisons are, however, limited by the contradictions between
the observations in different experiments, many of which have been
performed under different conditions.

\subsection{Summary of experimental results on NS3 helicase}

To compare our model to experiments on NS3 helicase, we would ideally
like to know the enzyme step size, the single-strand translocation
rate, and the double-strand unwinding rate---including information on
how it varies with NA sequence or applied force.  Interpretation of
experimental data on NS3 is complicated by differences in experiments
done by different research groups.  Some groups study the full-length
NS3 protein, including the helicase and protease domains
\cite{serebrov04,frick04c,beran06,dumont06,myong07}, while others
study the helicase domain only
\cite{preugschat96,levin99,levin02,levin03,levin04,frick04c,levin05}.
Moreover, genetically different versions of NS3 can have different
properties \cite{lam03b}.  The NS3 protein can also function in
different oligomeric states.  In bulk solution experiments,
full-length NS3 seems to function best as a dimer or higher-order
oligomer \cite{tackett05}, but single-molecule experiments can observe
unwinding by NS3 monomers \cite{dumont06,myong07}.  The helicase
domain NS3h appears not to form dimers in solution
\cite{levin99,kim98,porter98b}, but multiple copies of the protein can
bind to ssNA and unwind dsNA \cite{levin99}. In at least one
experiment, the kinetic parameters did not vary with the length of the
ss tail used to load NS3h, suggesting that the helicase mechanism may
not depend on whether the protein is a monomer or dimer
\cite{levin04}.

Contradictory claims have been made in the literature on the
qualitative description of NS3 helicase as well as on its quantitative
characteristics. First, we consider the empirical evidence for the
stepping pattern and the step size of NS3 helicase.  Recently a
detailed computational model of NS3, based on known crystal
structures, supported the idea of single-base ``inchworm'' motion
taken by NS3 monomers.  This model of Zheng et al.~proposes a major
protein conformational change which is triggered by ATP binding and is
coupled to forward motion of the helicase \cite{zheng07}.  Models
based on structural studies of NS3 have suggested single-base steps
\cite{yao97,kim98}. Similarly, structures of the distantly related
Hel308 helicase, which shows some structural similarities to NS3,
supports the idea of a ratchet-like mechanism during the ATP cycle
\cite{buttner07}.  However, most experimental efforts to determine the
step size don't support single-base steps.  Bulk kinetic experiments
have given a kinetic step size of 9-17 basepairs, depending on protein
form and unwinding substrate \cite{levin04,serebrov04,beran06}.
Single-molecule experiments on monomers of full-length NS3 have
suggested a step size of 11 basepairs with 3 basepair substeps
\cite{dumont06} or 3 basepairs with 1 basepair substeps
\cite{myong07}.  The most recent single-molecule work has proposed
that the fundamental step size is one basepair, with pauses occurring
less frequently as part of the ssNA bound to the helicase occasionally
``rips'' off \cite{myong07}.

Next we summarize the current estimates of ss translocation rate and
the speed of double-strand unwinding by NS3 helicase.  The maximum ss
translocation rate can be estimated from experiments that measure the
ATP hydrolysis rate. In one experiment, the NS3h rate of ATP
hydrolysis had a maximum $k_{cat}$ of 80 s$^{-1}$ in the presence of
the single-stranded oligo dU$_{18}$ \cite{preugschat96}.  Assuming
that during ss translocation the helicase hydrolyzes 1 ATP per step,
this measurement sets an upper bound on the ss translocation velocity
of 80 bases s$^{-1}$.  The double-strand unwinding velocity of NS3 has
been estimated from bulk and single-molecule experiments.  In one
single-turnover bulk kinetic study, the maximum unwinding rate of NS3h
was 2.7 bp s$^{-1}$ \cite{levin04}; similar results were found by
another group \cite{frick04c}.  Full-length NS3 may unwind at higher
velocities, up to 16.5 bp s$^{-1}$ \cite{serebrov04,frick04c}.  In
single-molecule experiments with applied force, full-length NS3
monomers unwind at force-independent rates of 50 bp s$^{-1}$
\cite{dumont06}. This relatively high velocity may be possible because
of the applied force that reduces the energetic cost of opening the
NA.  In single-molecule FRET experiments on full-length NS3 monomers
where no force is applied, an unwinding rate of $k\approx 0.9$
s$^{-1}$ was measured for one base pair substeps \cite{myong07}---a
value closer to the bulk value measured for NS3h.

Finally, we examine the experimental data to investigate whether the
unwinding by NS3 helicase is active or passive.  The dependence of the
unwinding rate on the base-pair binding free energy was measured both
in single-molecule and bulk experiments. In the work of Dumont et al.,
the RNA unwinding rate of full-length NS3 monomers was approximately
independent of applied force in the range 9-17 pN \cite{dumont06}. In
this experiment, the applied force was relatively high: the double
strand melted at a force of 20 pN. In single-molecule experiments
using a similar experimental setup, Cheng et al.~\cite{cheng07}
observed a significant effect of varying the RNA sequence on the NS3
unwinding rate. This observation of Cheng et al.~indicates that a
passive unwinding mechanism may not be adequate to explain the
behavior of full-length NS3 helicase.  Further, the apparent
contradiction between the observations of Cheng et al.~\cite{cheng07}
and Dumont et al.~\cite{dumont06} may be reconciled if we abandon the
simple physical picture in which the base-pair binding free energy can
be altered in a similar way by applied force or by changing the
sequence. Recent bulk measurements examined the effects of sequence
variation on the unwinding rate of NS3h \cite{donmez07}; this work is
discussed below where we compare our theoretical predictions to
experimental results.

In order to motivate our minimal model for the NS3 helicase, we now
discuss the affinity of NS3h to NA and its modulation during the ATP
hydrolysis cycle. Binding experiments on NS3h found that when the
helicase is bound to an ATP analogue, it binds to NA more weakly than
when not bound to ATP or ADP \cite{levin99,levin02}. The change in
binding free energy is approximately 6 $kT$ at room temperature (15 kJ
mol$^{-1}$) \cite{levin05}.  In addition, the affinity of NS3h for ADP
is low, so release of hydrolysis products is expected to be rapid
\cite{levin02}.  These observations are the basis of the proposed
flashing-ratchet mechanism of NS3h. (However, we note that another
work has found no dependence of NA binding on the ATP hydrolysis state
\cite{frick04c}; the source of this difference is unclear.)  
\begin{figure}
\begin{center}
\includegraphics[width=3.25in]{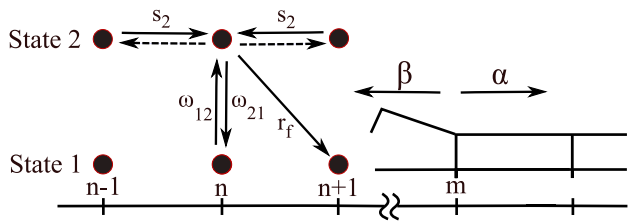}
\end{center}
\caption{Schematic of the simplified model that represents a flashing
  ratchet. The protein can exist in either of two chemical states
  (labeled 1 and 2) at each lattice site (labeled $n$). Sliding
  transitions (where $n$ changes but the state does not) occur only in
  state 2 at rate $s_{2}$.  Chemical transitions (where the state
  changes but $n$ does not) occur at rates $\omega_{12}$ (for the
  transition from 1 to 2) and $\omega_{21}$ (for the transition from 2
  to 1). A coupled transition (where both the state and $n$ change)
  occurs at rate $r_f$ (for the transition from 2 to 1 coupled to
  forward motion).  The nucleic acid single strand-double strand
  junction is at site $m$. The junction moves toward increasing $m$
  when the NA opens by one base (rate $\alpha$) and toward decreasing
  $m$ when the NA closes (rate $\beta$). }
\label{fig-ns3model}
\end{figure}

\subsection{Flashing-ratchet model of NS3 helicase}

Here we consider a special case of our model which corresponds to a
flashing ratchet mechanism.  Levin et al.~proposed that NS3 helicase
switches between two states: one tightly bound to the ssNA, the other
weakly bound \cite{levin03,levin05}. This scenario is referred to in
the physics literature as a flashing ratchet \cite{julicher97}.  When
applying the flashing ratchet scenario to NS3, the tightly bound state
is represented by a periodic sawtooth potential (with periodicity of
one ssNA base pair) and the weakly bound state is represented by a
uniform (weakly position-independent) potential \cite{levin05}. When
comparing to the flashing-ratchet scenario, we will consider state 1
to represent the strongly bound (S) state and state 2 the weakly bound
(W) state.  By comparing the theoretical predictions for this special
case of our model with the experimental data for NS3 helicase, we test
whether or not NS3 follows the flashing ratchet mechanism.

We assume that no sliding is possible in the tightly-bound state 1, so
$s_{1f}=s_{1b}=0$, and that the sliding is unbiased in state 2, so
$s_{2f}=s_{2b}=s_2$.  To connect with the flashing-ratchet scenario
and for simplicity, we assume that the rates $q_f=q_b=r_b=0$ (see
fig.~\ref{fig-ns3model}). With these assumptions, we find that the
rate of ss translocation is (from eqns.~\eqref{eq-kf} and \eqref{eq-kb})
\begin{equation}
v_{ss} = \omega_{12} \frac{r_f}{r_f+\omega_{12}+\omega_{21}},
  \label{vssns3}
  \end{equation}
and the rate of ds unwinding is
\begin{equation}
 v_u= \frac{\omega_{12}}{2} \frac{ (cr_f-(1-c)s_2)}
 {cr_f+\omega_{12} +\omega_{21}  }. \label{vns3}
\end{equation}

The excitation rate $\omega_{12}$ is associated with ATP binding, and
so is assumed proportional to ATP concentration. Therefore we write
$\omega_{12}=\omega_o \mbox{[ATP]}$. The rates $\omega_{21}$ and $r_f$
represent the relaxation from the weakly bound to the tightly bound
state that occurs after ATP hydrolysis, product release, and diffusion
in the weakly bound state. For a flashing ratchet, a high rate of
forward motion will occur when the positions of the energy barriers
and the time constants are such that forward movement (rate $r_f$) and
return to the same place after one cycle (rate $\omega_{21}$) occur
with equal probability. To match this optimal case, we therefore
assume that $\omega_{21}=r_f$. Further, we assume that the sliding
rate $s_2$ is small compared to the other rates; for concreteness we
will suppose $s_2 = \epsilon r_f$ with $\epsilon=0.1$ unless otherwise
stated. The velocities then become
\begin{eqnarray}
  v_{ss} &=& \frac{r_f \omega_o \mbox{[ATP]} }{\omega_o
    \mbox{[ATP]}+2r_f}, \label{vssl}\\ 
  v_u&=& \frac{(c-\epsilon(1-c))}{2} \frac{  r_f \omega_o \mbox{[ATP]}}
  {\omega_o \mbox{[ATP]} + (1+c) r_f  }. \label{vns3s} 
\end{eqnarray}

Both $v_{ss}$ and $v_u$ are consistent with the Michaelis-Menten
equation for enzyme kinetics, but with slightly different forms. Their
ratio is
\begin{equation}
  \frac{v_u}{v_{ss}} = \frac{(c-\epsilon(1-c))}{2} \frac{\omega_o
    \mbox{[ATP]}+2r_f}{\omega_o \mbox{[ATP]} + (1+c) r_f }.
\end{equation}
In other words, we predict that the ratio of the unwinding velocity to
the single-strand translocation velocity depends on ATP concentration.
If we average over sequence variation in DNA \cite{betterton03}, we
get the estimate $c=\alpha/\beta \approx 1/7$. For the purpose of
quantitative illustration of the variation of $\frac{v_u}{v_{ss}}$
with ATP concentration, let us assume $\epsilon=1/10$.  Then,
$\frac{v_u}{v_{ss}} \approx 0.029$ at high ATP concentration and
$\frac{v_u}{v_{ss}} \approx 0.05$ at low ATP concentration. This
suggests that the ratio of the unwinding velocity to the single-strand
translocation velocity could vary significantly with ATP
concentration---the change is almost a factor of 2 for this example.

Next, we estimate $v_u$ and $v_{ss}$ for NS3 helicase.  The
single-strand translocation and unwinding velocities are fully
determined by the parameters $c$, $r_f$, $\omega_o$, $\epsilon$, and
ATP concentration; we now extract estimates of $r_f$ and $\omega_o$
from experimental data.  In experiments at high ATP concentration and
in the presence of ssNA, NS3h shows a maximum ATP hydrolysis rate of
80 s$^{-1}$ \cite{preugschat96}. If we take this value as the limiting
ss-translocation rate and assume single base-pair steps, then $v_{ss}
= $ 80 nt s$^{-1}$ in the limit of high ATP concentration. Using this
estimate of $v_{ss}$ in eqn.~(\ref{vssl}), we get the estimate
$r_f=80$ s$^{-1}$. This, in turn, implies that at high ATP
concentration the unwinding velocity $v_u \approx 0.029 v_{ss} \approx
2.3 $ bp s$^{-1}$.  This value is comparable to the values of 2.7 bp
s$^{-1}$ \cite{levin04} found for NS3h and $0.9$ bp s$^{-1}$ found for
the one-bp substeps of full-length NS3 \cite{myong07}. We note that
the unwinding velocity $v_u \ll v_{ss}$, as should be expected for
this model which assumes a passive helicase mechanism.  Experiments
studying how NS3 ATPase activity \cite{preugschat96} and unwinding
\cite{dumont06} vary with ATP concentration found a similar Michaelis
constant $K_m \approx$ 90 $\mu$M. Using this value of $K_m$ in
eqn.~\eqref{vssl}, we estimate $\omega_o=2 r_f/K_m \approx 1.8 \
\mu$M$^{-1}$ s$^{-1}$.

\begin{figure}
\begin{center}
\includegraphics[width=3.25in]{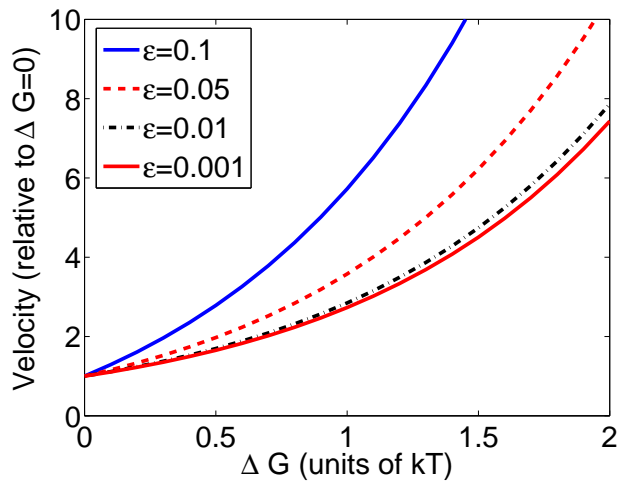}
\end{center}
\caption{Dependence of the unwinding velocity on the base-pair binding
  free energy. The reference state is a value $c=1/7$, which
  represents a sequence-averaged value for DNA. The additional
  destabilization energy $\Delta G$ (in units of $kT$) represents a
  free energy change that favors NA opening. When $\epsilon$
  increases, the dependence of the velocity on $\Delta G$ becomes more
  pronounced. However, decreasing $\epsilon$ can not flatten the curve
  indefinitely.}
\label{fig-forcedep}
\end{figure}

The only remaining unknown parameter is $\epsilon=s_2/r_f$, the ratio
of the sliding rate to the forward transition rate. A smaller value of
$\epsilon$ means that the sliding transitions in the weakly bound
state are less probable (see fig.~\ref{fig-ns3model}). A higher value
of $\epsilon$ means that sliding transitions in the weakly bound state
are more probable. This parameter has an important effect on the
dependence of the helicase velocity on the base-pair binding free
energy.

To study the effects of varying the base-pair binding free energy, we
focus on the limit of high ATP concentration. In this case, if $\Delta
G$ is the free energy of destabilization of base-pair binding, the
parameter $c=\alpha/\beta$ varies according to $c = c_o e^{\Delta G}$.
Therefore, at high ATP concentration, the unwinding velocity varies as
\begin{equation}
  \lim_{[ATP] \to \infty}v_u= \frac{(c_o(1+\epsilon)e^{\Delta G}-\epsilon)}{2} r_f.
\end{equation}
The unwinding velocity increases exponentially if the NA is
destabilized, as one would expect for a passive helicase.  However,
the precise shape of the curve of unwinding velocity versus $\Delta G$
depends on $\epsilon$. In the limit $\epsilon \to 0$, which physically
means no helicase sliding transitions occur in the weakly bound state,
the unwinding velocity varies with $\Delta G$ as a simple exponential:
\begin{equation}
  \lim_{[ATP] \to \infty, \epsilon \to 0 }v_u= \frac{c_o r_f }{2}
  e^{\Delta G}. 
\end{equation} 

As $\epsilon$ increases, the helicase can slide in the weakly bound
state. This allows more rapid unwinding by the helicase: when the dsNA
is destabilized, the ds base just ahead of the helicase has an
increased probability to be open. Rather than wait for the helicase
chemical transitions to move forward, the helicase can take advantage
of this increased junction open probability and slide forward. This
allows the steeper rate of increase of $v_u$ with $\Delta G$ seen in
fig.~\ref{fig-forcedep}.  This prediction is qualitatively consistent
with the result of Tackett et al.~\cite{tackett01}, who found that
full-length NS3 unwound double strands with higher melting
temperatures less efficiently. However, in the single-molecule
experiments of Dumont et al.~the unwinding rate of full-length NS3
monomers was practically independent of applied force in the range
9-17 pN \cite{dumont06}. This disagrees with the prediction of this
model, if the only effect of the applied force is to change the
binding free energy per base pair. However, this physical
interpretation is clearly not valid, because recent experiments from
the same lab find a significant variation in the RNA unwinding rate of
full-length NS3 with the variation of the base composition of the RNA
\cite{cheng07}. Reconciliation of the apparent contradictions in these
experimental observations is possible by assuming an active helicase
mechanism which, however, is not incorporated in the current version
of our model. Analyzing data from bulk experiments, Donmez et
al.~\cite{donmez07} claimed that the the variation of NS3h unwinding
velocity with base-pair binding free energy is inconsistent with a
passive helicase mechanism.  However, this conclusion is drawn from an
analysis based on a reported single-strand translocation velocity of
6.4 bases s$^{-1}$, which is much lower than the value of 80 bases
s$^{-1}$ mentioned above. A ss translocation rate of 80 bases
  s$^{-1}$ is an upper limit, assuming the helicase hydrolyzes 1 ATP
  per single-base step. If the helicase on average hydrolyzes $>1$ ATP
  per step, the ss translocation rate would be lower. A lower ss
  translocation rate would lead to an even larger disagreement between
  the passive helicase model we presented and the experimental data.
We believe that a conclusive comparison between our model of a
flashing-ratchet mechanism for NS3 helicase and the experimental data
is not possible because of the contradictory reports of experimental
studies.

\section{Conclusion}
\label{sec:disc}

In this paper we have developed a general model of unwinding of a
double-stranded nucleic acid molecule by a helicase motor. To capture
some of the key features of the helicase mechanochemical cycle, we
have modeled helicase switching between two chemical states. In this
model, the sites of a discrete lattice represent the positions of the
individual bases on the ssNA. At any spatial position, the helicase
can exist in either of the two allowed chemical states.  This model
should be generally applicable to helicases where one of the
transitions in the mechanochemical cycle is much slower than the other
transitions. In this work, we have considered only a passive helicase
mechanism---the helicase at the junction must wait for thermal
fluctuations to open the dsNA before it can advance. In future work,
it would be valuable to extend the model to include active
destabilization of the dsNA by the helicase.

To compare the model in detail to experimental data, we focused on a
special case which captures the flashing-ratchet mechanism proposed
for the NS3 helicase \cite{levin05}.  Solving the master equations for
this model at steady state, we have calculated the speed of unwinding
and the speed of single-strand translocation. The ratio of the
unwinding velocity to the ss translocation velocity varies with ATP
concentration as well as with the base-pair binding free energy.

Our comparison to experimental data on NS3 helicase suggests that the
model captures some features of the experiments. However, the
experimental literature on NS3 contains contradictory results.  This
may be a result of the different genetic variants, protein
truncations, oligomeric states, substrates, and buffer conditions used
by different laboratories. A set of detailed experiments by different
labs under consistent conditions may be important to fully understand
the unwinding mechanism of NS3 helicase.

\noindent {\bf Acknowledgments}: The authors thank Frank J\"ulicher
for several useful suggestions. DC also acknowledges support from the
Council of Scientific and Industrial Research (India) and the Visitors
Program of the Max-Planck Institute for Physics of Complex Systems,
Dresden (Germany).  MDB acknowledges support from the Alfred P. Sloan
Foundation and the Butcher Foundation.


\begin{thebibliography}{10}
\bibitem{note1}
We note that a two-state model for helicase translocation was also considered
  by Betterton and J\"ulicher \cite{betterton05a}, but the nature of the two
  states and the mechanism of translocation of the helicase are different from
  those developed here.
\end{thebibliography}

\end{document}